\documentclass[preprint,12pt,authoryear]{elsarticle}

\usepackage[T1]{fontenc}
\usepackage{graphicx}
\usepackage{afterpage}
\usepackage{orcidlink}
\usepackage{listings}

\newcommand{\ivxr}{{\sc{iv4XR}}}

\newcommand{\HIDE}[1]{}

\journal{Data \& Knowledge Engineering}

\begin{document}

\begin{frontmatter}

\author[upv]{Fernando Pastor Ricós}
\author[upv]{Beatriz Marín}
\author[utrecht]{I. S. W. B. Prasetya}
\author[upv,ou]{Tanja E. J. Vos}
\author[goodai]{Joseph Davidson}
\author[goodai]{Karel Hovorka}

\affiliation[upv]{organization={Universitat Politècnica de València},
            city={València},
            country={Spain}}

\affiliation[utrecht]{organization={Utrecht University},
            country={The Netherlands}}

\affiliation[ou]{organization={Open Universiteit},
            country={The Netherlands}}
    
\affiliation[goodai]{organization={GoodAI},
            city={Prague},
            country={Czechia}}

\title{Behavior Driven Development for 3D games}

\HIDE{
\begin{highlights}
\item The open-source iv4XR framework facilitates the creation of autonomous test agents for games.
\item Behavior-Driven-Development (BDD) supports human-readable and comprehensive test statements to verify functional aspects of games.
\item The iv4XR-BDD approach enables the testing of two 3D games through short and long interaction sequences.
\item The iv4XR-BDD approach has demonstrated to be a complement to traditional manual game testing, reducing the workload and improving effectiveness. 
\end{highlights}
}

\begin{abstract}
Computer 3D games are complex software environments that require novel testing processes to ensure high-quality standards. 
The Intelligent Verification/Validation for Extended Reality Based Systems (\ivxr{}) framework addresses this need by enabling the implementation of autonomous agents to automate game testing scenarios. 
This framework facilitates the automation of regression test cases for complex 3D games like Space Engineers. 
Nevertheless, the technical expertise required to define test scripts using \ivxr{} can constrain seamless collaboration between developers and testers. 
This paper reports how integrating a Behavior-Driven Development (BDD) approach with the \ivxr{} framework allows the industrial company behind Space Engineers to automate regression testing. 
The success of this industrial collaboration has inspired the \ivxr{} team to integrate the BDD approach to improve the automation of play-testing for the experimental 3D game LabRecruits. 
Furthermore, the \ivxr{} framework has been extended with tactical programming to enable the automation of long-play test scenarios in Space Engineers. 
These results underscore the versatility of the \ivxr{} framework in supporting diverse testing approaches while showcasing how BDD empowers users to create, manage, and execute automated game tests using comprehensive and human-readable statements. 
\end{abstract}


\end{frontmatter}

\noindent

\section{Introduction}
\label{section:introduction}
In the early 2020s, the video game industry captivated 2 billion global players and generated a revenue of 120 billion dollars \citep{bucchiarone2023games}. 
This growth raised the expectations of its audience, demanding high-quality products with engagement to long-term releases \citep{politowski2022towards}. 
As a result, the game industry has increasingly embraced agile development processes to facilitate rapid game feature enhancements. 
Nevertheless, there is a lack of test automation methods and tools for effective feature verification and efficient bug resolution \citep{politowski2021survey}. 
Consequently, game testing predominantly relies on manual efforts, with testers dedicating countless hours to ensure user interactions produce the intended responses in virtual scenarios. 

To tackle the limitation of game test automation solutions and provide a tool suitable for rapid software development processes, the Intelligent Verification/Validation for Extended Reality Based Systems project (\ivxr{} 2019-2022) initiated the development of the open-source \ivxr{} framework \citep{prasetya2022agent}. 
This framework aimed to implement autonomous testing \textit{agents} to help game industry stakeholders complement and enhance manual testing efforts \citep{prasetya2020aplib}. 

To address the regression testing challenges in the development process of the Space Engineers game, Keen Software House and GoodAI companies joined the \ivxr{} project. 
Its testers have devised a regression testing pipeline that comprises thousands of tests and requires dozens of hours of manual execution effort. 
The \ivxr{} framework appeared as a potential solution for seamlessly automating these regression tests into the development cycle. 
However, the technical expertise required to define test scripts within the \ivxr{} framework posed a comprehensibility challenge, limiting effective collaboration between testers and developers in creating and maintaining these automated test scripts. 

Behavior-Driven Development (BDD) is a software development process that fosters collaboration and communication between developers and testers \citep{smart2023bdd}. 
Although there is a scarcity of BDD research linked to industrial practices \citep{binamungu2023behaviour}, various studies have shown its increasing popularity and widespread use in diverse systems \citep{pereira2018behavior,bahaweres2020behavior}. 
Recognizing the advantages of BDD, the Space Engineers team embraced its adoption to streamline the \ivxr{} regression test automation. 

The successful integration of the BDD process with the \ivxr{} framework as an automated testing solution for Space Engineers has motivated the \ivxr{} project team to enhance the framework's capabilities. 
The experimental game LabRecruits, the primary 3D game demo of the \ivxr{} project, has been extended to support BDD for play-testing game levels. 
Additionally, the \ivxr{} game plugin used for automating regression tests in Space Engineers has been extended with tactical programming capabilities. 
This extension enables \ivxr{}-BDD \textit{agents} to perform intelligent navigational deliberations for long-play testing complex 3D scenarios in Space Engineers. 

{\bf Contribution.} This paper is an extension of a base version \citep{pastor2024industrial} presented at the \textit{18th International Conference on Research Challenges in Information Science (RCIS 2024)}, contributing the following:

\begin{itemize}

\item \textbf{Description of the testing practices} employed by the Space Engineers development companies. 

\item \textbf{Application of Behavior-Driven Development (BDD) to extend the \ivxr{} game test automation framework}, introducing automated agent-based testing into the Space Engineers regression testing pipeline. 

\item \textbf{Automation of an essential subset of the regression test suite}, providing the Space Engineers development companies with a reliable game automation approach. 

\end{itemize}

\noindent
The extension offers the following new contributions with respect to the base paper:

\begin{itemize}

\item \textbf{Enhanced support for play-testing in complex 3D scenarios}, achieved by incorporating tactical and goal-based programming into the \ivxr{}-BDD framework. While the previous regression test automation solution focused on short sequences of interactions, play-tests are usually lengthy, stretching over hundreds or even thousands of interactions.

\item \textbf{Extended technical explanation} of how BDD is integrated into the \ivxr{} framework, including a discussion of the practical challenges encountered during development and validation. 

\item \textbf{Broader validation} of the \ivxr{}-BDD solution through its application to two different 3D games: LabRecruits and Space Engineers. 

\end{itemize}

These contributions are valuable for game development researchers and practitioners as they offer insights into Space Engineers' game-testing methodology and \ivxr{} framework capabilities. 
The paper discusses the advantages and challenges of using automated approaches and highlights the potential integration of autonomous \textit{agents} through the \ivxr{} framework to improve automated test execution. 
Additionally, the paper provides guidance for practitioners who want to leverage the BDD process for effective and efficient automated execution of game-testing scenarios. 
 
The paper is structured as follows. 
Section \ref{section:related.work} presents the related work. 
Section \ref{section:iv4XR} describes game-testing automation challenges, the \ivxr{} framework as a solution, and the LabRecruits game. 
Section \ref{section:SE_game} introduces the Space Engineers game, the game testing practices, and the solutions and challenges of the \ivxr{} Space Engineers-plugin. 
Section \ref{section:BDD.testing} presents the benefits of the BDD testing process and its practical usage as an autonomous BDD \textit{agent}. 
Section \ref{section:industrial.bdd.se} describes the industrial application of BDD with Space Engineers. 
Section \ref{section:bdd.long.play.tests} provides insights about the extension of the \ivxr{}-BDD approach for long-play test scenarios with LabRecruits and Space Engineers games. 
Section \ref{section:conslusions.future.work} summarizes the conclusions and future work. 

\section{Related work}
\label{section:related.work}
In the last decades, there have been significant test automation advances for traditional desktop \citep{pezze2018automatic}, web \citep{garcia2020survey}, and mobile \citep{nie2023systematic} Graphical User Interface (GUI) systems \citep{rodriguez202130}. 
However, the field of game systems still lacks well-established and widely adopted methodologies to automate the execution of complex 3D games. 

One interesting solution is the VRTest framework \citep{wang2022vrtest}, which provides an interface to integrate different testing techniques using rotation, movement, and click-trigger events in Virtual Reality (VR) scenes. 
However, this framework requires further work to support wide types of events and be evaluated with software projects not based on Unity. 

In recent years, Machine Learning (ML) has been researched for game-testing automation.
ICARUS framework \citep{pfau2017automated} employs Reinforcement Learning (RL) to train agents that complete linear adventure games. 
Wuji's approach \citep{zheng2019wuji} uses Deep Reinforcement Learning (DRL) to train exploration agents that accomplish mission objectives.  
\cite{wu2020regression} evaluates multiple RL rewards for divergent behavior detection in two versions of a commercial game task. 
\cite{gordillo2021improving} apply curiosity-driven RL to identify areas that stuck players.  
\cite{deexploiting} combine metamorphic tests with RL techniques to train agents capable of detecting collisions and camera behavior failures. 
\cite{ariyurek2022playtesting} use RL to train agents that simulate different personas to discover alternative playstyles. 
\cite{sestini2022automated} use curiosity and imitation RL to train agents that explore goal-destination areas while uncovering collision bugs and glitches. 

These studies showcase the capability of ML techniques to automate game play-testing through trained agents. 
However, most studies are more concerned with the performance of the ML models rather than developing methodologies that assist companies in automating the verification of test case goals and oracles used to determine if a test passes or fails \citep{politowski2022towards}. 
This emphasis on ML often leads stakeholders to perceive game-testing automation approaches as complex solutions since using ML requires, e.g., good knowledge of different models, learning algorithms, their hyperparameters, and the technology stack involved. In all the ML studies mentioned above, test agents are trained to do a general goal, e.g., to complete the target game or to explore the game world as much as possible. Moreover, agents trained in this way are not necessarily suitable for testing specific scenarios. 
These complexities and limitations of ML raise concerns about its practical usage and maintenance. 
Therefore, it is essential to research methodologies that empower non-technical testers to design, create, and maintain human-readable tests. 

The application of BDD in game testing remains an underexplored area. 
Among the few studies found, one notable example is the RiverGame framework \citep{paduraru2022rivergame}, which adopts a BDD approach to engage non-technical stakeholders in describing desired game features and evaluating their usability and correctness. 
The RiverGame study primarily focuses on evaluating the accuracy of their visual recognition techniques with demo applications and their voice detection rate approach using an industrial game. 
In contrast, our work presents a BDD approach with the \ivxr{} framework that enables the automation of short regression tests and long-play test scenarios, focusing on verifying the functional aspects of the game under test. 
 
In summary, while ML techniques have demonstrated exploratory test effectiveness, these can be perceived as complex solutions for testers who require a scenario-based testing approach that guides specific actions to assess concrete functionalities. 
Thus, we emphasize the importance of creating solutions that seamlessly facilitate the versatile integration of various tester-friendly methods. 
To address this, we took two main steps: (i) we used the \ivxr{} framework to establish a connection with the Space Engineers game environment, allowing the observation of internal game objects and executing game actions; and (ii) we extended the \ivxr{} framework with the BDD testing process approach for designing test scenarios in natural language for LabRecruits and Space Engineers 3D games. 

\section{Game-testing automation}
\label{section:iv4XR}
To address the challenges of game test automation, the \ivxr{} framework provides a solution that enables test \textit{agents} to connect to, observe, and interact with objects within game environments. 

\subsection{Challenges in game-testing automation}
Record-and-replay and other scripted GUI testing methods have shown fragility and unreliability, attributed to unpredictable GUI system behaviors and evolving GUI changes made between versions \citep{hammoudi2016record,coppola2018mobile}. 
Despite this, the GUI testing field has seen considerable progress. 
For example, accessing the web DOM data to generate reliable test scripts \citep{nguyen2021generating}. 
However, this progress has limitations for 3D games due to the inability of existing tools to obtain the required game system information. 

Automated tools for game systems rely on recording keyboard and mouse inputs, lacking access to internal game data such as position, orientation, and object properties. 
Thus, successfully reproducing recorded test actions becomes challenging, as the traversed states may exhibit slight variations in each replay. 
Unlike traditional GUI testing, this is not due to a test script being vulnerable to GUI modifications but rather because the test lacks the necessary resilience to account for the inherent dynamics and non-determinism of the game. 
To achieve effective game test automation, it is crucial to have access to the game object's positions for precise navigation and to the object's properties for test oracles. 

\subsection{iv4XR framework for game-testing automation}
The \ivxr{} framework consists of a plugin architecture that includes a set of Java classes that streamline connection, information retrieval, and interaction with game objects. 
This framework supports the integration of test \textit{agents} capable of connecting with and controlling playable game characters \citep{prasetya2022agent}. 

To use the \ivxr{} framework, the interested stakeholders must develop a dedicated game-plugin that adheres to the framework architecture. 
This plugin must contain a set of \textit{Environment}, \textit{Entity}, and \textit{Commands} interfaces that enable seamless integration of the \textit{agent} for connecting with game scenarios, observing existing game entities and their properties, interacting with the game system by executing the defined actions, and apply test oracles. 

\begin{figure}[!ht]
  \centering
  \includegraphics[width=\linewidth]{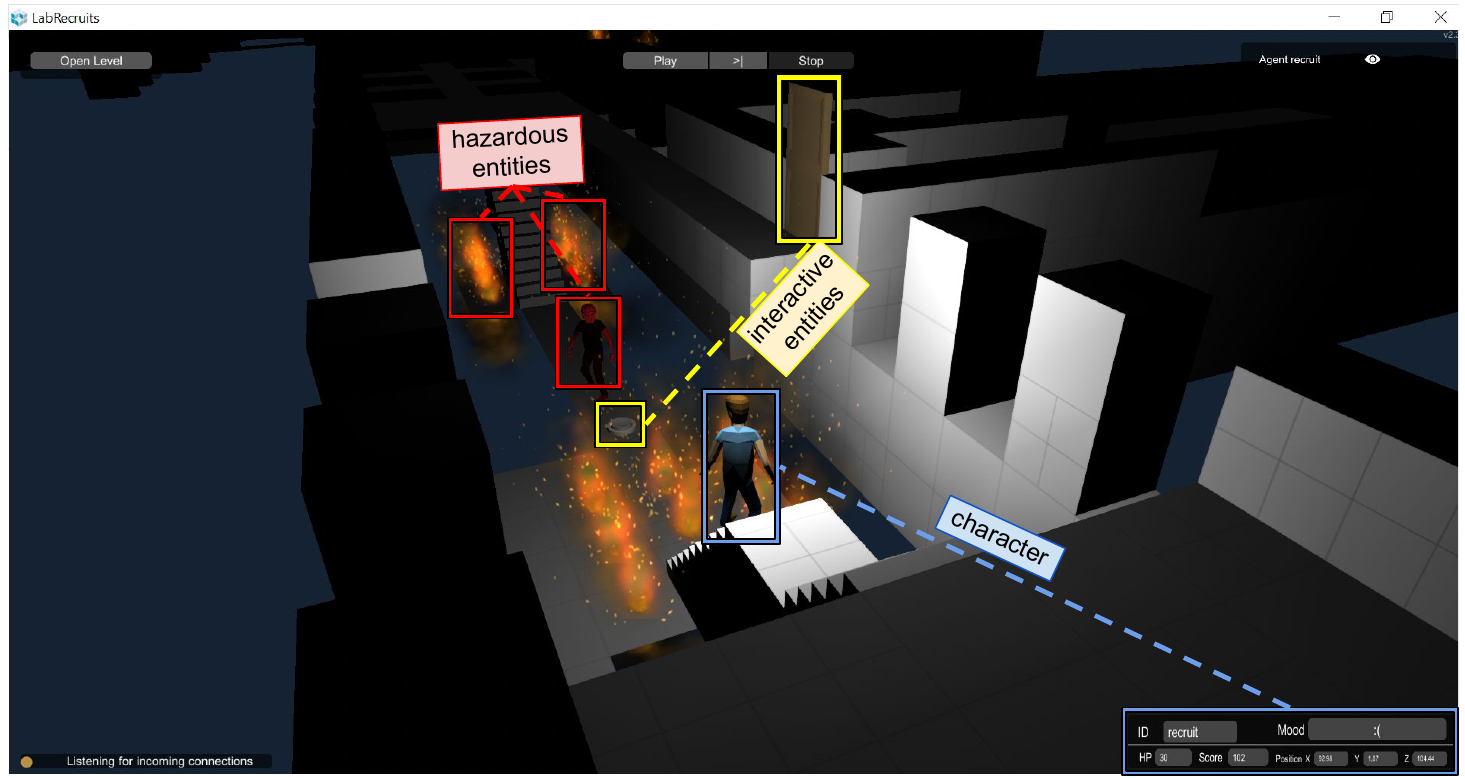}
  \caption{LabRecruits game objects}
  \label{fig:LR_game}
\end{figure}

LabRecruits \footnote{\url{https://github.com/iv4xr-project/labrecruits}} is an experimental 3D game provided by the \ivxr{} framework as a demo example for integrating AI \textit{agents} to test game environments. 
The \ivxr{} framework includes the LabRecruits-plugin \footnote{\url{https://github.com/iv4xr-project/iv4xrDemo}} as an example to enable access to the internal data and functions of this 3D experimental game. 
In this game, test \textit{agents} can explore the environment to reach a flag objective while interacting with buttons and doors and avoiding hazardous entities. 
Figure \ref{fig:LR_game} shows a screenshot of the game, featuring interactive entities such as one button and one door, a hazardous monster, various fire hazards, and the playable game character. 

\begin{figure}[!ht]
  \centering
  \includegraphics[width=\linewidth]{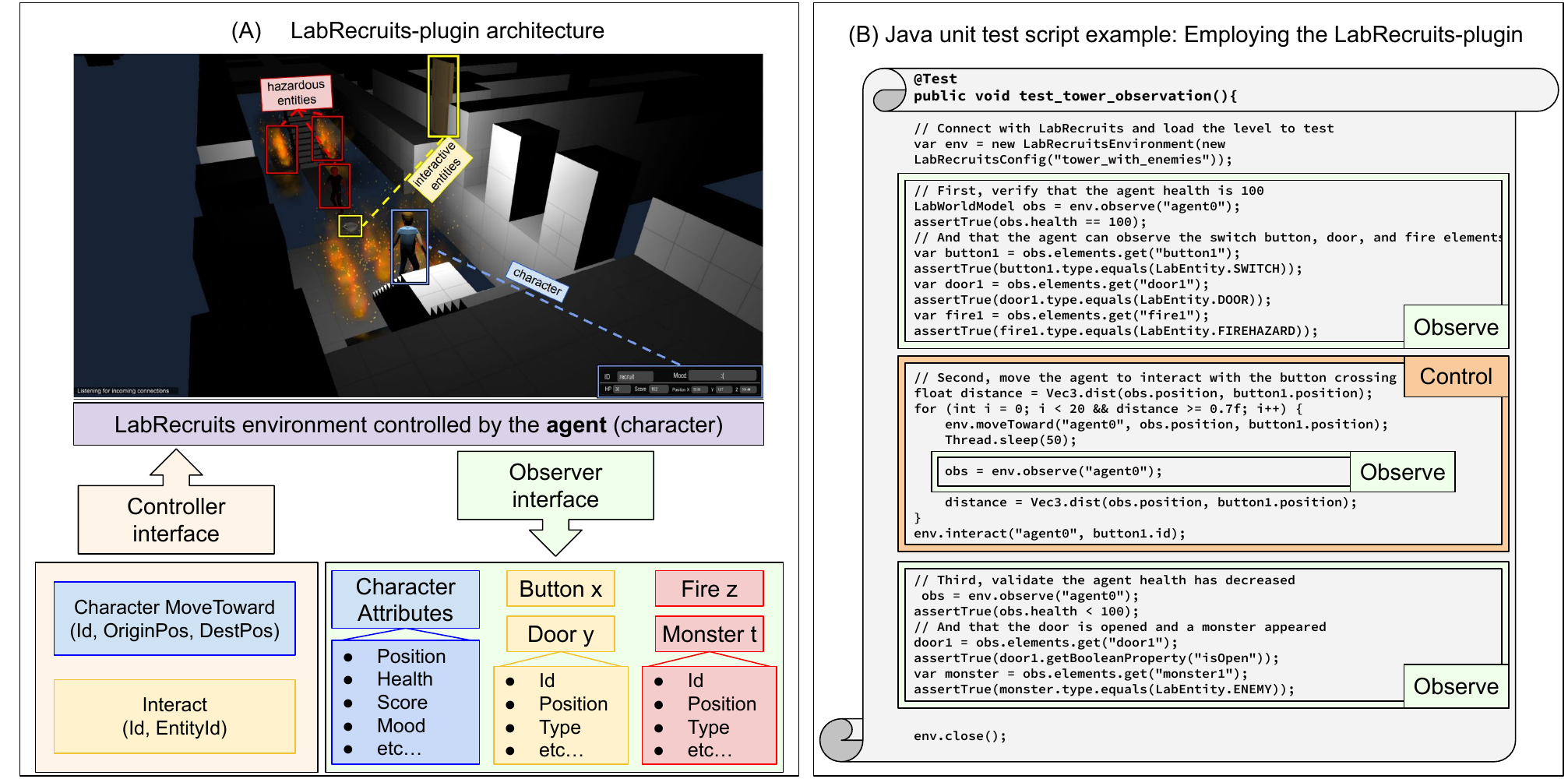}
  \caption{Overview of the LabRecruits-plugin and Java unit test script example}
  \label{fig:LR_plugin_script}
\end{figure}

Once the LabRecruits-plugin enables the integration of an \textit{agent} for connecting with and controlling a playable character in LabRecruits, \ivxr{} stakeholders can implement Java unit test scripts to apply test oracles.  
Figure \ref{fig:LR_plugin_script} illustrates an example of the LabRecruits-plugin architecture (A) and a Java unit test script that employs the LabRecruits interfaces (B). 

The \textit{agent} can gather information about nearby game objects by observing the LabRecruits environment. 
Since the character is always present, the \textit{agent} can observe its position and attributes continuously. 
Nonetheless, the observation of entities such as buttons, doors, and hazardous elements depends on the proximity of the \textit{agent} and the presence of obstructive walls. 
For example, if the \textit{agent} is near a button, it will be able to observe it; however, being far away or separated by a wall would prevent the \textit{agent} from observing the button. 
 
In the Java unit test script example (B), the \textit{agent} first observes that his health is complete and confirms the presence of specific button, door, and fire game elements. 
The \textit{agent} is then controlled to move toward the button, moving through the fire, and continues until it is close enough to interact with the button. 
After interacting with the button, the \textit{agent} observes that his health has decreased due to fire damage, that the button interaction has opened the door, and that a monster enemy has appeared in the game environment.\looseness=-1 

Motivated by the potential of the \ivxr{} framework as a test automation solution for the game Space Engineers, Keen Software House and GoodAI companies decided to investigate the integration of this approach into their development cycle. 

\section{Space Engineers game}
\label{section:SE_game}
Space Engineers is an industrial 3D sandbox game that allows users to assume the role of an astronaut, a playable game character capable of interacting with realistic open-world scenarios. 
Since its initial alpha release in 2013 until late 2023, the game has continuously evolved through around 600 game updates and ongoing maintenance for feature updates and bug fixes. 

\subsection{Game objects}
\begin{figure}[!ht]
  \centering
  \includegraphics[width=\linewidth]{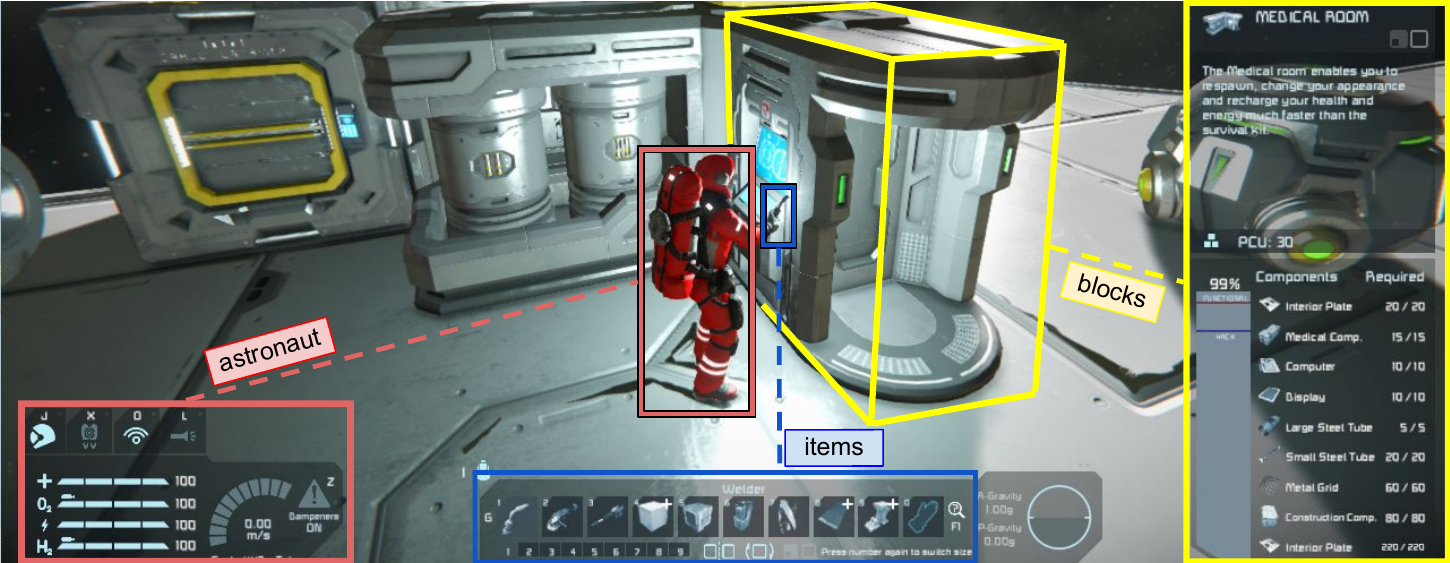}
  \caption{Space Engineers game objects}
  \label{fig:SE_game}
\end{figure}

In Space Engineers, users have the freedom to explore diverse planets and asteroids, where they can design and build space stations and spaceships for spatial exploration. 
Resource management is a vital game aspect, requiring players to organize and gather materials to thrive in space environments. 
Figure \ref{fig:SE_game} shows the main game objects: the \textit{astronaut}, the \textit{items}, and the \textit{blocks}. 

The \textit{astronaut} character owns various attributes, including energy, hydrogen, oxygen, and health, which players must manage to ensure their survival. 
To explore space scenarios, the \textit{astronaut} is equipped with movement, rotation, and jet-pack flying capabilities. 
Additionally, the \textit{astronaut} can utilize \textit{items} to build \textit{blocks}, with these \textit{items} falling into two main categories: tools and components. 

The welder and the grinder are the primary tools. 
The welder empowers players to construct \textit{blocks}, provided the \textit{astronaut} has the necessary components. 
Conversely, the grinder enables the destruction of \textit{blocks} to retrieve components. 

Each game \textit{block} resides in a specific position and orientation and has properties that represent their type, integrity, volumetric physics, mass, inertia, and velocity. 
Certain types of \textit{blocks} possess functional capabilities, such as gravity and energy generation or restoring the astronaut's health. 
The construction of multiple interconnected \textit{blocks} forms a \textit{grid}, where different \textit{blocks}, such as a medical room \textit{block} and an oxygen generator \textit{block}, can be connected via conveyor \textit{blocks}. 
The wide variety of \textit{blocks} exhibits dynamic functional interoperability with other \textit{grid} \textit{blocks} (e.g., a medical room connected to an oxygen generator can also restore the astronaut's oxygen). 
Consequently, Space Engineers is an extensive and complex game that poses significant testing challenges due to the vast range of blocks and their diverse and intricate interactions. 

\subsection{Current testing practices}
\begin{figure}[!ht]
  \centering
  \includegraphics[width=\linewidth]{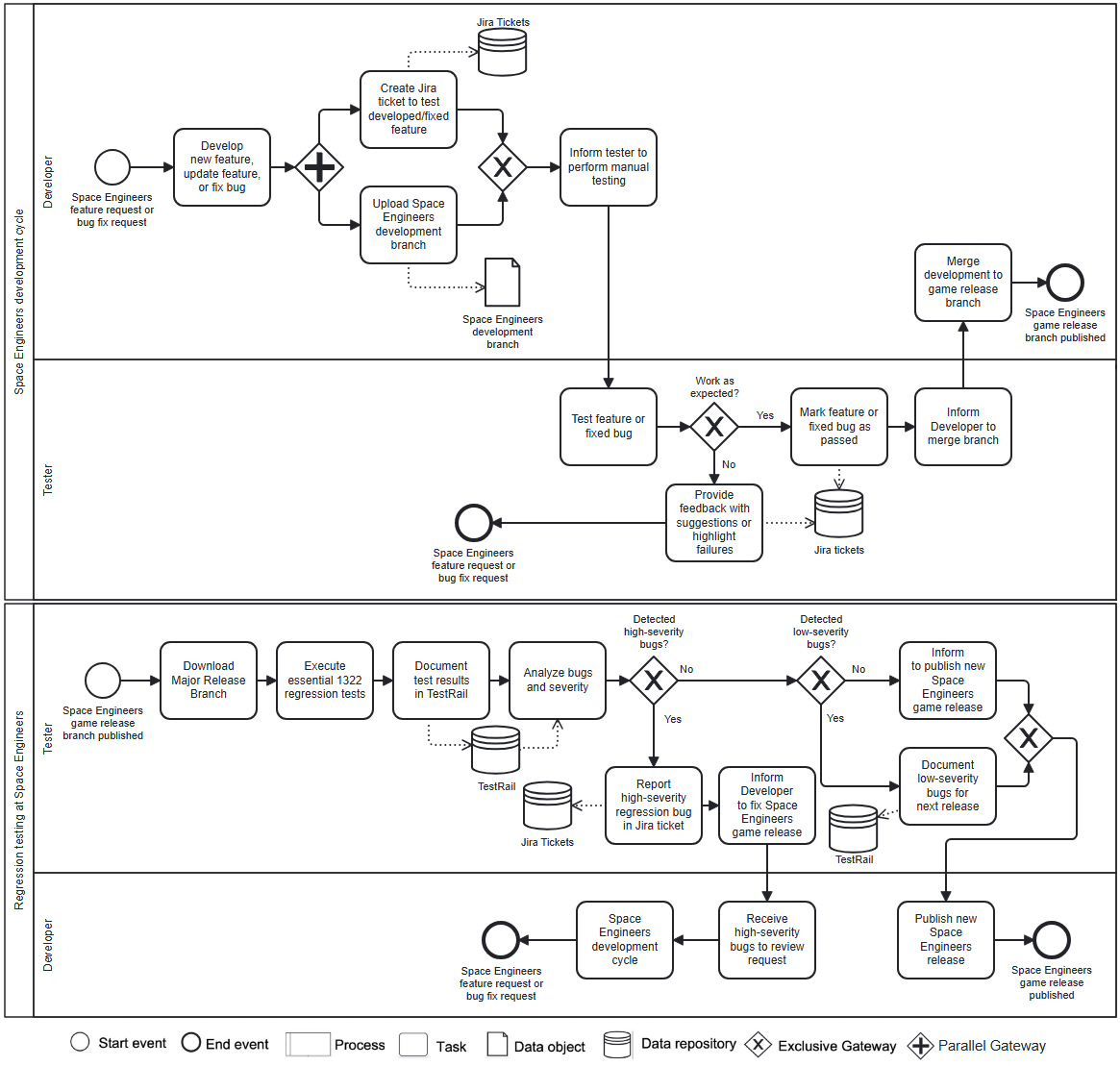}
  \caption{Space Engineers development cycle diagram}
  \label{fig:SE_dev_diagram}
\end{figure}

During a Space Engineers \textit{game release}, a development cycle begins with the team of developers working on changing the game to add, update, or fix features (Figure \ref{fig:SE_dev_diagram}). 
As developers finalize these feature changes, they open Jira tickets\footnote{\url{https://www.atlassian.com/software/jira}} to point testers to the development branch that contains features that require testing. 
Testers then process these Jira tickets and {\em manually} test the newly implemented, updated, or fixed features. 
This development cycle takes about 3 or 4 months, depending on the required changes of the new game version. 

Once developers have implemented all the required changes, testers assess whether the changed features have reached a stable state. 
If testers discover bugs in some of the changed features, they report the unstable features to the developers, who continue improving the development.  
After addressing any identified bug and confirming the stability of the development branch, testers proceed with executing a regression test suite. 

This suite comprises 1322 tests, encompassing game features related to the astronaut's movements, attributes, items usage, interactions with blocks, as well as graphic or sound aspects. 
Testers manually execute the 1322 tests and document the results in TestRail\footnote{\url{https://www.gurock.com/testrail/}}, a management system that supports the organization of the testing process. 
Upon completing the regression test suite, testers analyze the results to determine if all tests passed without detecting any bugs. 
However, if any bugs are found during testing, testers assess each bug's severity. 

If testers determine that a bug is of high severity, they open a Jira ticket, requiring developers to address it and fix the bug before proceeding with the game release. 
However, if the bug is classified as low-severity, the development branch can be published as a new \textit{game release}, with the understanding that the low-severity bugs will be fixed in subsequent releases. 

The regression testing phase described above is a time-consuming task that requires approximately 96 hours of manual effort. 
Due to the substantial manual time investment, integrating the regression test suite into the development cycle of a \textit{game release} is not feasible. 
Instead, the regression testing is executed at the end of the cycle, typically occurring every 3 or 4 months. 
This delay in regression testing may result in possible bugs that could have been detected and fixed in the iterative development process but are now postponed for several months, potentially impacting the final steps of the \textit{game release} process. 

The proficiency and perception of human testers are essential to ensuring the accurate verification of astronaut, item, and block functionalities, as well as graphic and sound aspects. 
However, the Space Engineers team recognized the possibility of automating a subset of functional regression tests that are repetitive and time-consuming. 
By automating this functional regression subset, nightly builds can seamlessly integrate into the development cycle. 
This integration will enable the early detection of possible regression bugs, significantly improving the overall effectiveness and efficiency of the \textit{game release} process. 

\subsection{iv4XR technical solution and comprehensibility challenge}
Before joining the \ivxr{} project, the Space Engineers team evaluated the use of record-and-replay and visual recognition techniques to automate the execution of regression tests. 
However, they found a significant challenge as the recorded scripts proved to be unreliable in nearly all tests, even executed in the same scenario and game version. 
Reproducing the recorded astronaut movements and interactions did not yield the expected outcome, highlighting the lack of reliability in the recorded scripts. 

\begin{figure}[!ht]
  \centering
  \includegraphics[width=\linewidth]{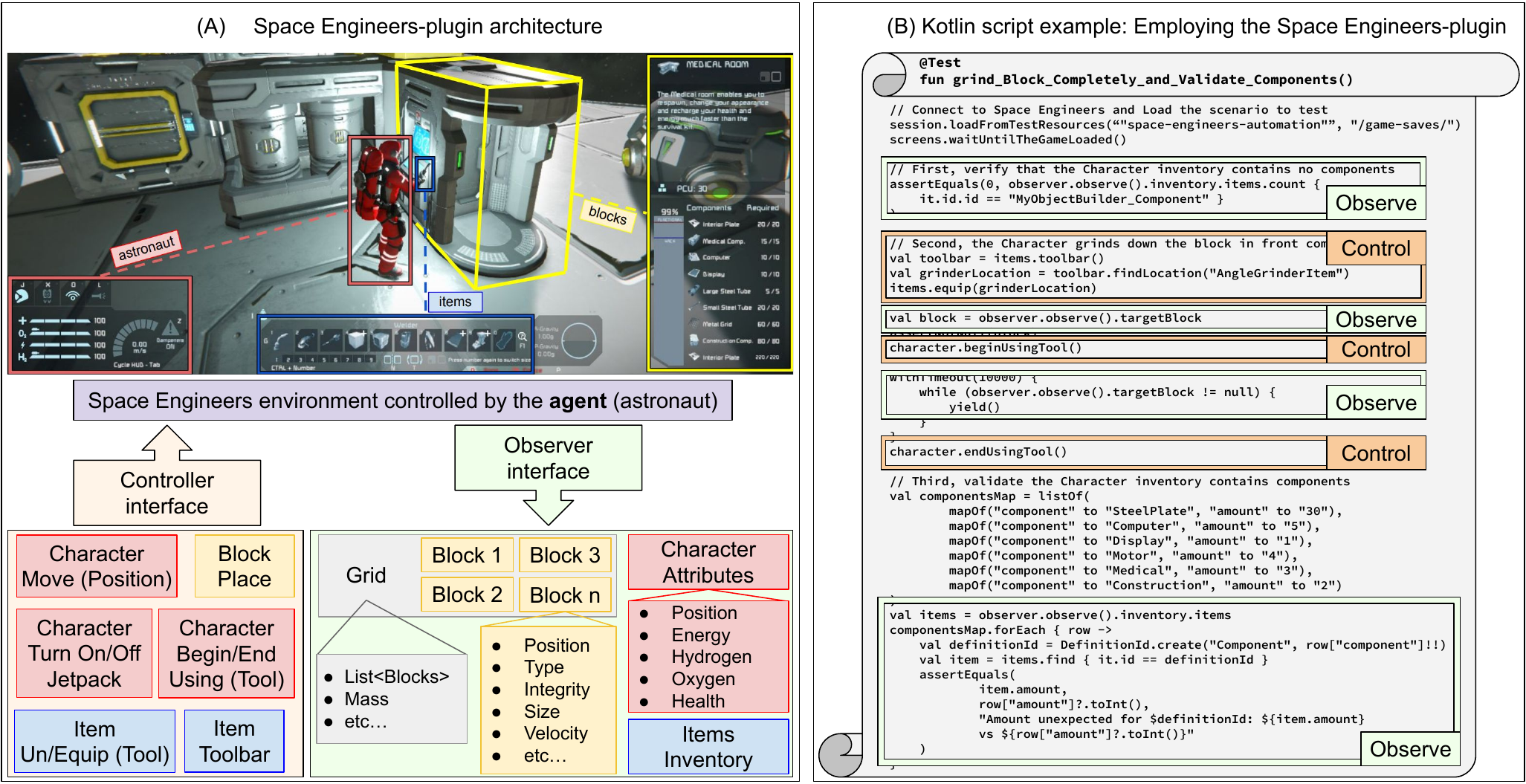}
  \caption{Overview of the Space Engineers-plugin and Kotlin test script example}
  \label{fig:SE_plugin_script}
\end{figure}

Recognizing the limitations of record-and-replay tools in game automation, the Space Engineers team opted to incorporate the \ivxr{} framework to create reliable regression tests that integrate directly with the game's internal data.
To implement this functionality, we developed a dedicated Space Engineers-plugin in Kotlin\footnote{\url{https://github.com/iv4xr-project/iv4xr-se-plugin}}. 
This plugin enables the integration of an autonomous \textit{agent} capable of connecting with and controlling the astronaut playable character within Space Engineers. 
Figure \ref{fig:SE_plugin_script} shows an overview of the plugin's architecture (A) and a Kotlin test script example that employs the plugin's interfaces (B). 

The \textit{agent} gathers information about the surrounding game objects by \textit{observing} the Space Engineers scenario. 
Since the astronaut character is always present, the \textit{agent} can continuously monitor its own character attributes and inventory. 
However, block observation depends on the proximity of the \textit{agent} to different grids and block entities. 
For instance, if the \textit{agent} is near a construction, it will be able to observe all of the blocks, whereas being far away would not allow it to observe them. 

In the Kotlin test script example (B), observing the game objects involves identifying the inventory and the target block aimed by the \textit{agent}.  
The inventory is observed initially to ensure it contains no items. 
At the end of the test script, it is observed again to validate that a specific amount of items can be found. 
On the other hand, observing the target block allows for checking its existence or whether it has been destroyed. 

To ensure a reliable astronaut \textit{agent} control, the Space Engineers-plugin incorporates \textit{controller} classes that interact with internal game functions. 
These classes translate actions into executable game commands empowering the \textit{agent} to maneuver the astronaut to move and rotate to specific vector positions, equip or unequip various items such as tools, utilize equipped items, and place equipped blocks.  
In the Kotlin script example (B), first, the items toolbar is controlled to find and equip the desired tool for the character. 
Once the tool is equipped, the character is controlled to begin using the tool until the target block is destroyed. 
When the target block is destroyed, the character ends up using the tool. 

The \ivxr{} Space Engineers-plugin addresses the technical side and enables the creation of reliable test scripts to observe and manipulate the game environment through the \ivxr{} \textit{agent}. 
This feature enables the automation of the regression test suite. 
However, during this industrial collaboration, we encountered another challenge: the comprehensibility of the test scripts. 

Game development involves team members with varying levels of expertise, including management, sound designers, artists, programmers, and testers. 
Even among programmers, there is a diverse range of technical skills related to specific game packages. 
As a result, not all team members possess the technical knowledge needed to create test scripts using the iv4XR Space Engineers-plugin architecture. 

To foster comprehensive collaboration between members involved in the iv4XR Space Engineers-plugin and game testers for creating regression tests and to promote the creation of reusable tests, we decided to adopt the principles of Behavior-Driven Development (BDD) \citep{smart2023bdd}. 
As explained in subsequent sections, this approach empowers game testers to design, generate, and maintain evolving test scenarios using natural language statements. 

\section{Behavior-Driven-Development for test automation}
\label{section:BDD.testing}
To successfully implement the BDD process in industrial practice, it was essential to integrate a comprehensive suite of tools and techniques on top of the Space Engineers-plugin. 
This integration empowers testers to abstractly design and automate test script execution, freeing them from the need to delve into the technical intricacies of the plugin. 

\subsection{Given-When-Then theoretical structure}
In a typical test scenario aimed at validating a game feature, the technical implementation involves the creation of a test script class with the following steps: 
\begin{enumerate}
    \item \textbf{Load Game Scenario:} The test script should programmatically load a specific game scenario containing the desired state objects and properties that require testing. This step sets the initial conditions for the test. 
    \item \textbf{Execute Actions:} The next step involves running a specific set of actions that will alter the desired game object. These actions simulate user interactions or in-game events that trigger changes in the system. 
    \item \textbf{Validate Results:} After executing the actions, the test script should validate that the game object and its properties have responded adequately to the actions. This verification ensures the game features work as intended. 
\end{enumerate}
To implement these steps in a Space Engineers test scenario, testers can create a test script in Kotlin utilizing specific functions from the Space Engineers-plugin. 
However, crafting this Kotlin script requires technical knowledge of the existing plugin functions and valid game-plugin variables, which can pose challenges for non-technical testers in designing and maintaining test scripts. 

The Given-When-Then (GWT) structure \citep{north2006introducing} is a standard format widely used in the BDD process. 
It enables the abstraction and description of system component functionality and expected behavior in a human-readable language: 
\begin{itemize}
    \item \textit{Given} statement defines the Space Engineers level scenario and the game state that the test aims to validate. For instance, it can involve loading a specific level with the desired block to be tested. 
    \item \textit{When} statement enables to control the astronaut and execute a predefined set of actions that alter the Space Engineer's game objects. For example, the test instructs the astronaut to equip a grinder tool and destroy a block.  
    \item \textit{Then} statement verifies that the game object's properties have changed as expected, according to the specifications of the game feature being tested. For instance, it could check if the astronaut's inventory contains a specific amount of components after the block is destroyed.
\end{itemize}

\begin{figure}[!ht]
  \centering
  \includegraphics[width=\linewidth]{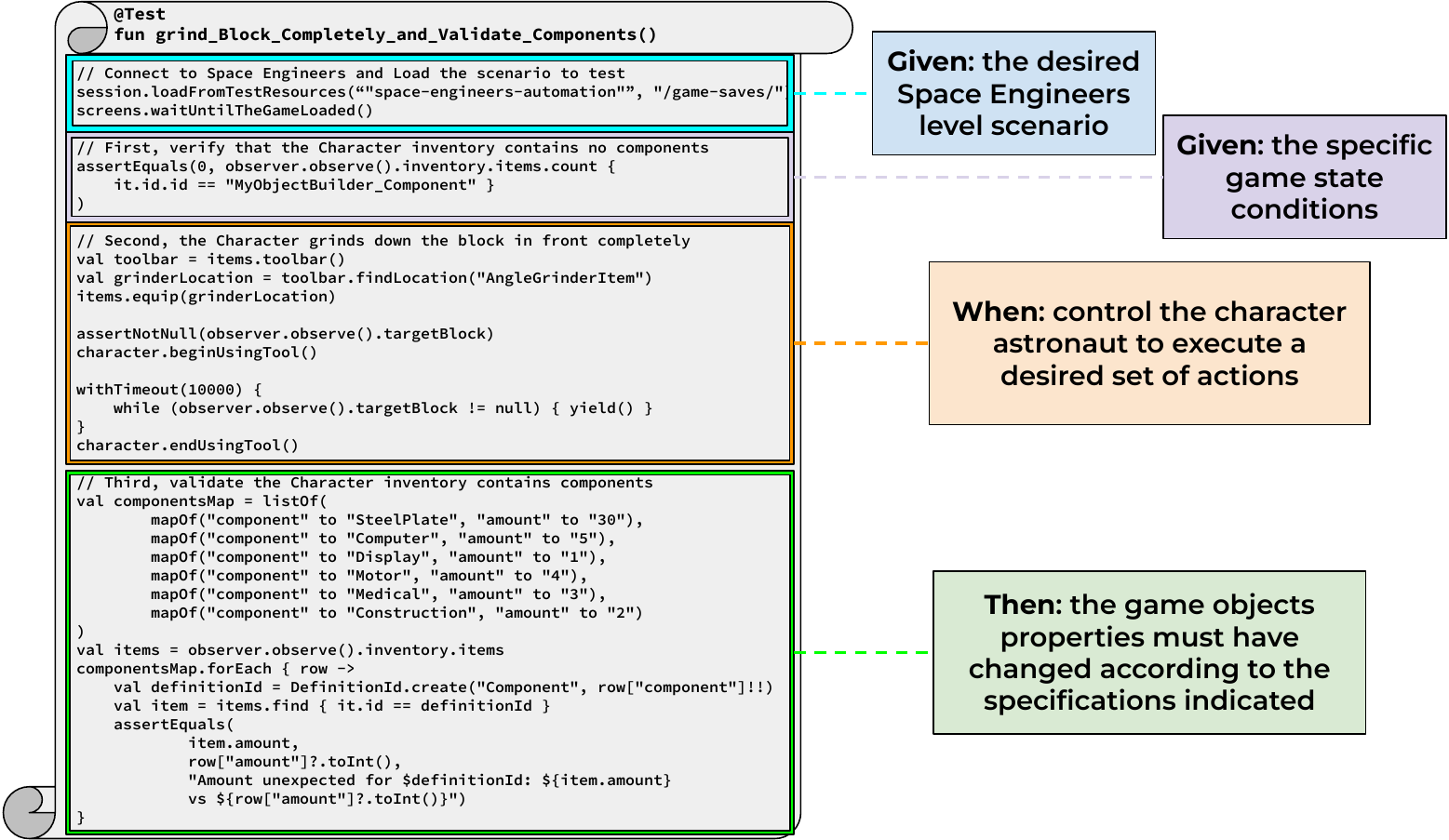}
  \caption{Space Engineers Kotlin test script with GWT structure}
  \label{fig:SE_kotlin_script}
\end{figure}

Figure \ref{fig:SE_kotlin_script} illustrates how the Space Engineers test script written in Kotlin can be structured in a GWT structure. 
By adopting this structure in Space Engineers, testers can design and maintain regression test scenarios straightforwardly. 

\subsection{BDD in practice with Cucumber}
Cucumber \citep{dees2013cucumber} is an open-source tool that facilitates the practical implementation of the GWT structure in the context of a BDD process. 
This tool has been integrated as a bridge between human-readable GWT statements and the technical Space Engineers-plugin functions that interact with the game. 

\begin{figure}[!ht]
  \centering
  \includegraphics[width=\linewidth]{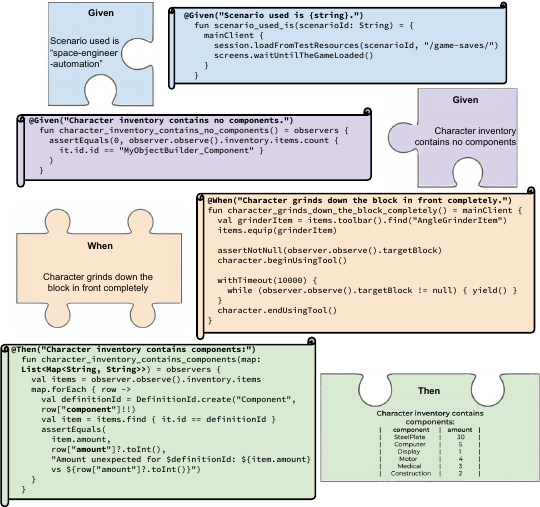}
  \caption{Cucumber mapping of Space Engineers-plugin functions with GWT statements}
  \label{fig:SE_cucumber_functions}
\end{figure}

During the implementation with Cucumber, Kotlin functions are created to encapsulate the necessary invocations, aligning with the objectives specified in the \textit{Given}, \textit{When}, or \textit{Then} statements. 
Figure \ref{fig:SE_cucumber_functions} shows how the original Kotlin script can be disassembled into functions to load game scenarios, execute actions, or validate game properties. 
Each corresponds to a specific GWT statement. 

Once these functions are implemented, they can be seamlessly aggregated and invoked using Cucumber's human-readable language. 
This approach offers Space Engineer's testers an intuitive way to create, maintain, and execute test scenarios within the Space Engineers game environment. 

\begin{figure}[!ht]
  \centering
  \includegraphics[width=\linewidth]{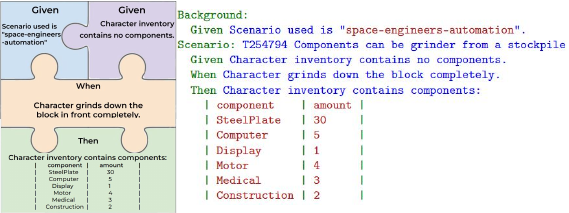}
  \caption{Space Engineers BDD test scenario to validate astronaut inventory components}
  \label{fig:SE_cucumber_puzzle}
\end{figure}

In Figure \ref{fig:SE_cucumber_puzzle}, we present a human-readable test script resulting from the combination of the BDD process with Cucumber. 
This script represents a concrete Space Engineers regression test scenario that non-technical stakeholders can craft to validate the grinding of a block to collect components by the astronaut. 
In the subsequent section, we delve into the association between these test scenarios and specific game objects, such as T254794, providing a comprehensive understanding of how these scenarios are applied in practical testing situations. 

\section{Industrial application of iv4XR-BDD for Space Engineers}
\label{section:industrial.bdd.se}
Using the integration of BDD with the iv4XR Space Engineers-plugin, the Space Engineers team began to design the necessary game levels and the automation of regression test scenarios. 

\subsection{Space Engineers pre-designed game level for testing}
\label{section:results.bdd}
\begin{figure}[!ht]
  \centering
  \includegraphics[width=\linewidth]{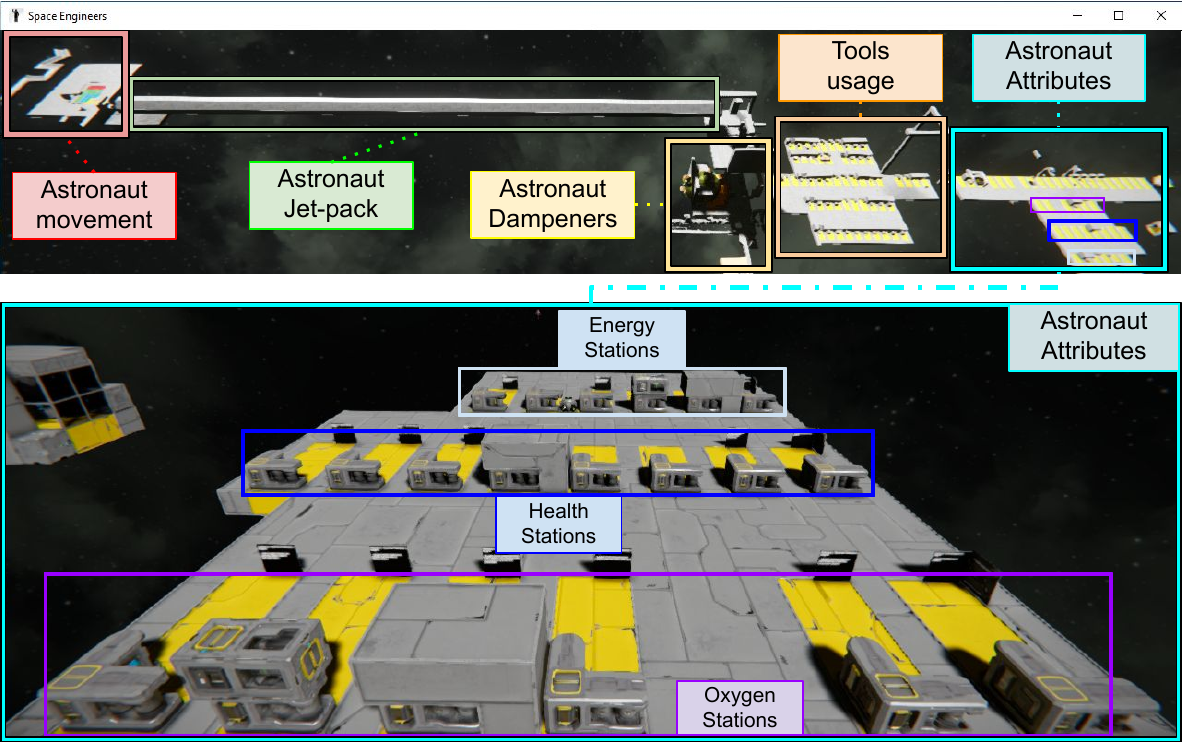}
  \caption{Space Engineers pre-designed game level for testing}
  \label{fig:SE_pre_degined_level}
\end{figure}

Executing BDD regression tests in a sandbox open-world game like Space Engineers poses an additional challenge. 
Our goal is to keep astronaut actions (\textit{When}) and validation oracles (\textit{Then}) as minimalist as possible. 
To do this, testers have pre-designed a game level (see Figure \ref{fig:SE_pre_degined_level}) that can be loaded using the \textit{Given} statement. 
This level, divided into sections and stations, provides all the necessary blocks and items required by the astronaut in the diverse BDD test scripts. 

Each section groups together generic features, such as astronaut movements, attributes, jet-pack, dampeners, and block grinding. 
In each section, multiple stations are set up with custom-arranged blocks designed to test specific functionalities. 
For example, a section is dedicated to validating the astronaut's health, oxygen, and energy attributes. 
In one medical station equipped with an oxygen tank, we validate the agent's oxygen replenishment. 
At another medical station without an oxygen tank, we validate that the agent cannot replenish his oxygen. 

Each BDD test scenario is linked to the specific station that helps validate the intended functionality. 
To transit between test scenarios, the Space Engineers team utilizes a process that spawns the testing \textit{agent} at the designated station. 

\subsection{Regression testing with automated BDD test scenarios}
The Space Engineers team decided to initiate the automation process with a subset of 236 tests from the total of 1322 regression tests for the following reasons: 
\begin{enumerate}
    \item \textbf{Essential Features:} The selected subset comprises essential features for validating the astronaut's movements and attributes. These functionalities play a significant role in ensuring an immersive player experience. 
    \item \textbf{Plugin Classes:} The current state of the plugin covers all the necessary functions required for automating the control of the astronaut's movement and interactions with blocks using tools and items. 
    \item \textbf{Oracle Validation:} It is feasible to integrate a reliable Oracle validation, which involves observing the state of the game, including assessments of the astronaut's position and attributes. Test scenarios that involved validating game graphics or sounds currently present challenges for validation. 
    \item \textbf{Manual Effort Time:} While executing each individual test is not time-consuming, the entire subset requires a considerable amount of time. 
\end{enumerate}
 
The Space Engineers team has successfully automated a subset of 236 regression tests using the BDD \textit{agent}. 
In contrast to previous experiences, the BDD approach has demonstrated \textit{reliability} in controlling the astronaut and applying validation oracles. 
This automation effort offers an opportunity for seamless integration of these 236 regression tests into nightly builds, marking an enhancement from the prior practice of conducting them solely at the end of the 3 to 4-month development cycle. 
Instead, the BDD \textit{agent} can complement the daily verification work that testers do manually. 
The Space Engineers team estimates that this automation initiative could be equivalent to saving approximately 17 hours of manual effort each time the regression tests have to be executed. 

Each BDD test scenario automatically generates a TestRail entry indicating whether the test passed or failed. 
If a test fails, the automated process generates a detailed report containing textual and visual information about the failed step. 
This seamless integration with TestRail has proven effective by successfully identifying a regression testing bug related to the engagement of the astronaut's magnetic boots. 
This success motivates further integration in nightly builds to streamline the regression test suite during the \textit{game release} development cycle. 

\subsection{Threats to validity}
This section presents threats that could affect the validity of the Space Engineers' industrial results \citep{wohlin2012experimentation,ralph2018construct}. 

\textbf{Content validity} 
While the number of successful automated regression tests can be quantitatively assessed, evaluating the satisfaction of the Space Engineers team is more subjective and difficult to measure from an analytical standpoint. 
In future work, we plan to continue automating regression tests using the BDD \textit{agent} and evaluate team satisfaction through qualitative feedback. 

\textbf{Internal validity}
The selection of regression tests for automation relied on the expertise of the Space Engineers team. 
However, this experience could introduce selection bias and impact the representativeness of chosen regression tests. 
To mitigate this threat, the team fostered a collaborative environment where developers and testers with high expertise in the game led the selection process. 

\textbf{External validity} 
This industrial study with Space Engineers serves as a testing reference for a complex sandbox game. 
However, the aforementioned regression test automation mainly covers short sequences of interactions, rather than play-tests in complex 3D scenarios, which are usually lengthy executions, stretching over hundreds or even thousands of interactions. 
Additionally, it is essential to acknowledge that games significantly vary in implementation and design. 
Therefore, to assess the generalizability of the \ivxr{} and BDD approaches, it is recommended to evaluate their effectiveness and applicability across a wider range of games. 

\textbf{Conclusion validity}
The equivalence of saving 17 hours of manual effort resulting from automating the initial 236 regression tests provides a compelling advantage. 
However, it is important to acknowledge the absence of a controlled experiment, which limits the generalizability and statistical robustness of the conclusions. 
Future research should prioritize conducting other empirical evaluations to validate and reinforce the observed benefits of automated regression tests. 

\subsection{Lessons learned}
The current collaboration to address the lack of automated regression test execution in an industrial environment provided us with valuable lessons learned: 

\textbf{Lesson one: Do not underestimate the complexity of automating oracles when testing 3D sandbox games}
Multimedia aspects like visual textures, UI text displays, pixel particles, animations, lighting, and sounds are as crucial as correct block and item functionalities. 
However, the complexity of 3D perspectives, dynamism, and non-determinism (e.g., weather systems or NPC behaviors) make multimedia oracle verification a complex and unreliable endeavor that also introduces subjectivity in the form of diverse user experiences. 
For example, certain regression tests that have not been automated yet demand human testers to move the astronaut on the ground of planets or spatial structures or fly with the jetpack while continuously checking animations and textures. 
The complexity in verifying these animations and textures underscores the difficulties in automating human checks as technical oracles. 
Achieving this automation would require integrating an approach capable of recording a running video of the astronaut's 3D perspectives and accurately verifying the dynamic multimedia aspects. 

The physics in 3D sandbox games, marked by non-deterministic, randomness, and unpredictability, complicates precise test oracle automation. 
For instance, Space Engineers testers are developing self-crafted environments that automate the execution of collision of rockets and vehicles while measuring range variables with in-game sensor entities. 
However, due to the non-deterministic physics, human validation is necessary to ensure the accuracy of the game's responses. 

In multiplayer games, challenges intensify when verifying oracles from dedicated server and client player perspectives. 
The extensive game functionalities and multimedia aspects must be verified in dynamic multiplayer environments, where it is necessary to confirm the correct synchronization of game events (e.g., a bug provoked that astronaut faction changes were accurately synced from the client but were not triggering the desired changes when invoking from the server itself).  
Moreover, these complexities extend to validating network protocols and cross-platform compatibility (e.g., a recent issue resulted in console players being unexpectedly disconnected after minutes of playing on computer servers). 

\textbf{Lesson two: Automation efforts must concentrate on supporting human testers rather than aiming for their outright replacement}
During the selection of regression test cases intended to integrate the BDD automation process, we learned a lesson that exceeds technical challenges: the value of human tacit knowledge \citep{walker2017tacit}. 

Humans possess the capacity for comprehension, intuition, and accumulated understanding that is sometimes challenging, if not impossible, to express or formalize in technical scripts. 
Furthermore, this knowledge and experience is subjective, varying according to each individual perception and personal beliefs. 
Attempting to automate this human tacit knowledge as test scripts is an unrealistic goal that should be avoided when integrating a software automation process. 

In the case of Space Engineers, which contains thousands of regression tests, automating astronaut controls and validating individual game properties through integrated test oracles is essential for reducing manual effort. 
However, the wide variety of game objects that can be used in dynamic and highly customizable environments makes it challenging to anticipate every possible scenario. 
For instance, although game blocks exhibited correct functionality when tested individually, a user discovered that removing a block connecting a self-constructed station to an asteroid caused the entire station to move inside the asteroid. 

An expert human tester with a broad tacit knowledge background in testing Space Engineers can intuitively understand the combination of game blocks and physics causing the problem. Leveraging this tacit knowledge, they can notify developers of potential causes and apply their expertise to future developments affecting these specific buggy game features. 

Therefore, the primary objective should not be focused on replacing human testers with artificially embedded tacit knowledge in a testing procedure. Instead, the focus should be on complementing and supporting human testers to streamline their work, reduce their workload, and provide them with information that enhances their productivity. 

\textbf{Lesson three: The dynamic nature of the industry requires flexibility and adaptability in academic-industry collaborations}  
In long-term projects, the industry is more likely to change team members by incorporating new personnel or reallocating existing members to projects deemed of higher priority. 
Consequently, in academic-industry collaborations, flexibility and adaptability are essential. 
Proactive anticipation of potential changes in team compositions, along with the recognition that industry members may face varied future workloads, enables smoother transitions and enhances the resilience of the collaboration. 
This entails strategic foresight, planning, and meticulous documentation for effective knowledge transfer. 
Within this collaborative environment, future industry members or external open-source contributors can quickly assimilate into the project, fostering a fluid and robust long-term environment. 

\section{Using iv4XR-BDD for long-play test scenarios}
\label{section:bdd.long.play.tests}
The integration of the \ivxr{}-BDD approach with the technical controller and observer interfaces of LabRecruits and Space Engineers enables human testers to define behavior-driven tests using natural language statements. This solution has allowed the Space Engineers team to automate a subset of their regression tests, focusing on unit and short interaction scenarios, verifying astronaut properties \citep{pastor2024industrial}. 

However, the previous \ivxr{}-BDD solution, which operates at the level of low-level game commands and observations, is mainly suited for short interaction sequences. For example, a short regression test might spawn the Space Engineer astronaut near a block with all required inventory materials, simply to verify that constructing a specific block is possible. 

While effective for testing isolated game functionalities, this solution does not address the challenges of supporting long test executions that aim to simulate complex user behavior and navigational decision-making in dynamic 3D game environments. For instance, a more realistic play-test scenario might involve spawning the astronaut in an open-world environment, requiring them to explore, locate, and grind multiple construction materials before placing and welding a block, simulating the decisions a real player would make over a long play session. 

To overcome these limitations, it is necessary to extend the \ivxr{}-BDD approach to support play-testing. Play-tests are longer, more complex testing scenarios designed to simulate longer gameplay sessions. Rather than verifying isolated or unit-level functionalities, play-testing involves hundreds or even thousands of interactions that more closely simulate human player behavior. These longer test executions demand intelligent deliberation and adaptive navigation, particularly for managing 3D movement, pathfinding, and goal-oriented actions within complex game scenarios. 

\subsection{Tactical and Goal iv4XR programming for play-testing}
In addition to game command interactions, the \ivxr{} framework supports the use of structured \textit{Tactics} and \textit{Goals}. 

Tactics consist of a prioritized sequence of tasks that an \ivxr{} \textit{agent} executes across deliberation cycles until a specified test goal is achieved. Each task serves as a fallback: if the command interactions of one task fail or are not applicable in the current game state, the agent automatically proceeds to the next, enabling it to adapt its behavior dynamically to the environment. 

For example, Listing \ref{tactic.navigate} presents a tactic that aims to navigate to a specific game entity and is composed of four sequential prioritized tasks: (1) attempt to plan a path to the entity using the current navigation graph; (2) if a path is found, try to follow it and navigate to the entity’s position; (3) if the entity is not visible in the current observed navigation graph, explore the environment to discover new reachable areas; and (4) if none of the above steps succeed (e.g., the entity is unreachable), abort the tactic to avoid infinite retries.

\begin{figure}[!ht]
\begin{lstlisting}[language=Java, label={tactic.navigate}, caption={A tactic that aims to navigate to a game entity}, captionpos=b, frame=single, numbers=none, basicstyle=\footnotesize]
Tactic navigateTo(String entityId) {
    return FIRSTof(
        planPath(entityId),
        navigatePathTo(entityId),
        explore(),
        ABORT()
    );
}
\end{lstlisting}
\end{figure}

Similarly, Listing \ref{tactic.interact} shows a simpler tactic that aims to interact with a game entity. The agent first attempts the interaction, and if that fails (e.g., the entity is out of range), it defaults to aborting.

\begin{figure}[!ht]
\begin{lstlisting}[language=Java, label={tactic.interact}, caption={A tactic that aims to interact with a game entity}, captionpos=b, frame=single, numbers=none, basicstyle=\footnotesize]
Tactic interactWith(String entityId) {
    return FIRSTof(
        interact(entityId),
        ABORT()
    );
}
\end{lstlisting}
\end{figure}

These tactics define how an \ivxr{} \textit{agent} should act to achieve a specific test goal. A goal encapsulates a condition, typically expressed as a predicate over the agent’s belief state, and associates it with a tactic intended to fulfill that condition. Goals can be used as individuals or combined into \textit{goal structures}, which allow for expressing sequential goal prioritization. 

For example, Listing \ref{goal.entity.interact} illustrates a sequential goal structure that combines two individual goals: (1) the first goal ensures that the agent navigates near a specified entity; and (2) once that goal condition is met, the second goal makes the agent interact with the entity. 

\begin{figure}[!ht]
\begin{lstlisting}[language=Java, label={goal.entity.interact}, caption={A sequential goal structure for navigation and interaction with an entity}, captionpos=b, frame=single, numbers=none, basicstyle=\footnotesize]
GoalStructure entityInteracted(String entityId) {
    // First goal: navigate near the entity
    var goalNavigate = goal("Navigate near the entity")
        .toSolve((BeliefState BS) -> position(entityId) < 0.3)
        .withTactic(navigateTo(entityId))
        .lift();

    // Second goal: interact with the entity
    var goalInteract = goal("Interact with the entity")
        .toSolve((BeliefState BS) -> true)
        .withTactic(interactWith(entityId))
        .lift();

    return SEQ(goalNavigate, goalInteract);
}
\end{lstlisting}
\end{figure}

This tactical programming approach enables the integration of \textit{agents} capable of making intelligent deliberation decisions to adaptively cope with a changing environment, non-determinism, and hazardous entities \citep{prasetya2020aplib}. Thanks to this adaptivity, tactical tests are very robust \citep{ShirAplibRobust21}. Moreover, in combination with a navigation layer, these deliberation cycles allow \textit{agents} to autonomously perform long navigation and exploration tactics to discover and reach game entities, reducing the need to programmatically issue numerous individual movement commands \citep{prasetya2022agent,shirzadehhajimahmood2025automated}. 

\begin{figure}[!ht]
  \centering
  \includegraphics[width=\linewidth]{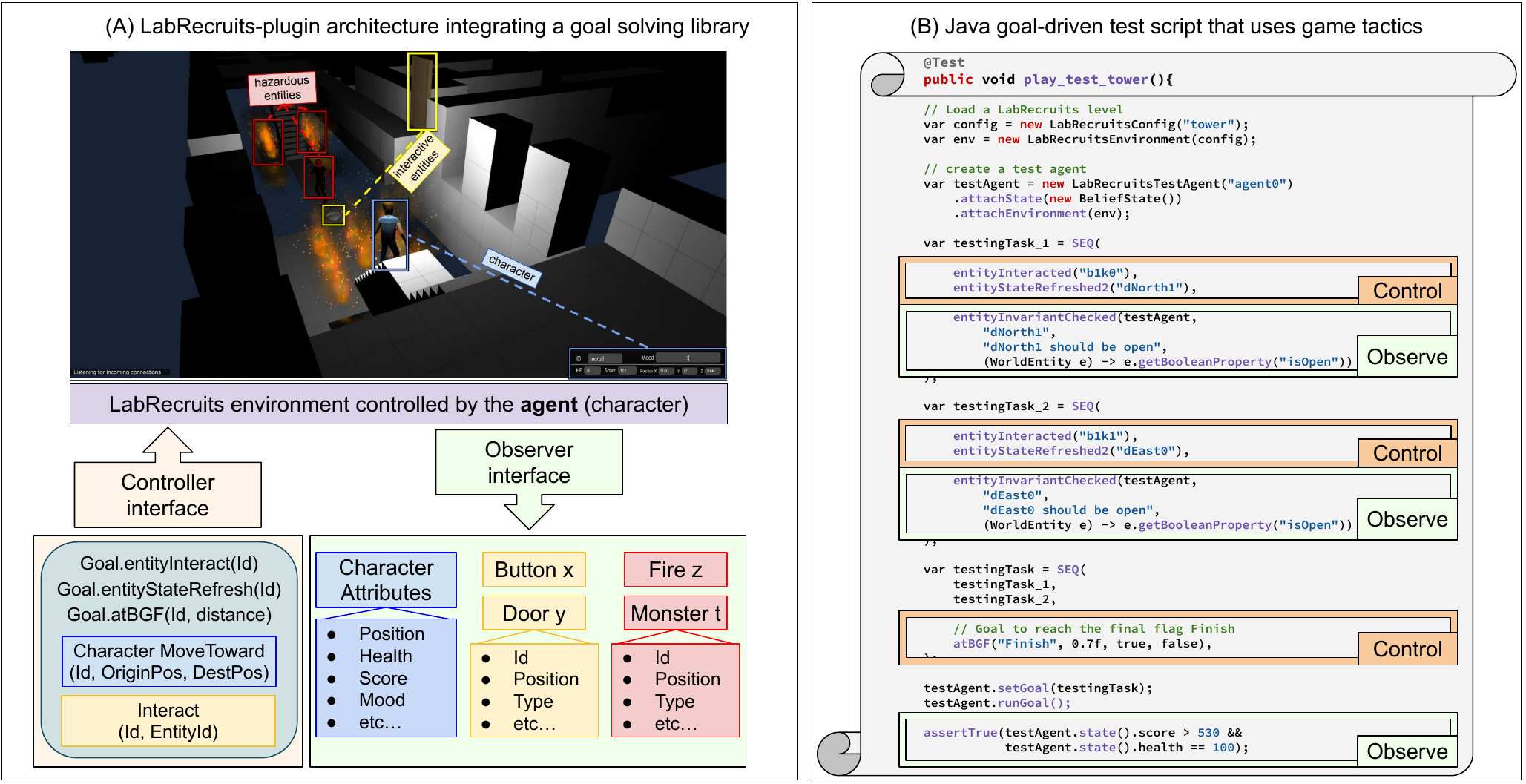}
  \caption{Overview of tactical programming in the LabRecruits-plugin with a Java goal-driven test script that uses LabRecruits game tactics}
  \label{fig:LR_plugin_tactic}
\end{figure}

Figure \ref{fig:LR_plugin_tactic} illustrates the incorporation of the high-level \textit{Goal} library within the controller interfaces of the LabRecruits-plugin architecture (A) and a Java goal-driven test script that employs these tactics and goals to perform long-play tests of LabRecruits levels (B). 
In the Java goal-driven test script example (B), the \textit{agent} is instructed to follow a scenario specified by a series of goals that can be solved with game tactics. 
In the case of LabRecruits, these tactics involve discovering and interacting with specific button entities, observing the status of particular doors after interacting with the buttons and applying invariant test oracle goals to verify that the doors have been opened. 
At the end of these tactical executions, the \textit{agent} aims to reach the finish flag to complete the goal-driven test script objective and assesses that all goals were accomplished by maintaining 100\% of its health and achieving at least 530 score points. 

\subsection{Long-play iv4XR-BDD test scenarios for LabRecruits}
During the industrial experience with Space Engineers, members of the \ivxr{} project recognized the value of adopting the BDD approach to improve the comprehensibility of \ivxr{} tests for non-technical stakeholders. 
While the goal-driven test script for long-play testing utilizes a high-level library that abstracts much of the low-level technical interfaces of \ivxr{}, the combination of tactics and goals can still pose challenges for stakeholders unfamiliar with the framework. 
Based on insights gained from this experience, we decided to extend the LabRecruits-plugin demo to incorporate the BDD approach. 
This extension aims to improve the comprehensibility of LabRecruits game test scenarios, making it easier for non-technical users to create, maintain, and understand goal-driven test scripts. 

\begin{figure}[!ht]
  \centering
  \includegraphics[width=\linewidth]{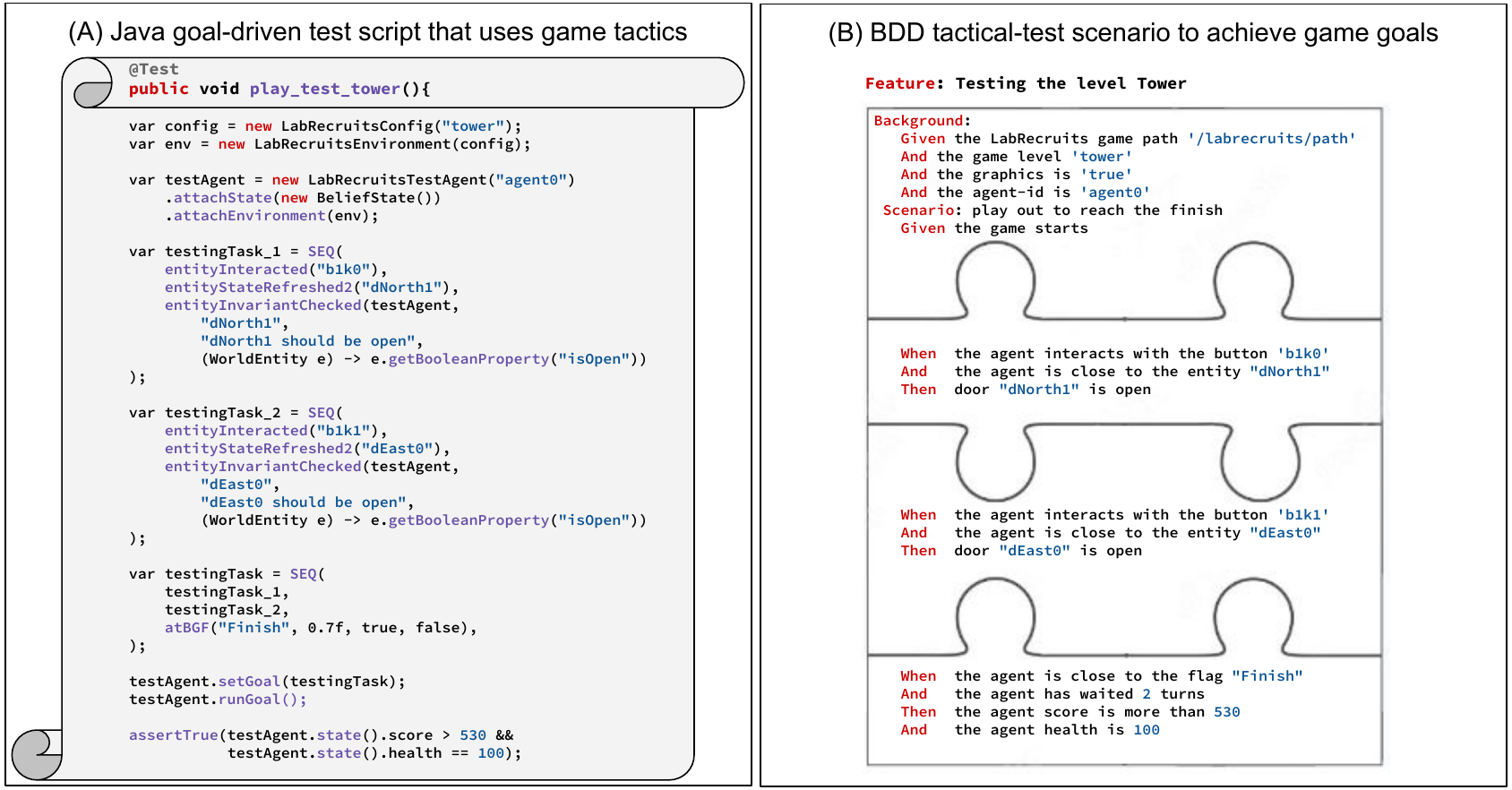}
  \caption{LabRecruits long-play iv4XR-BDD test scenario}
  \label{fig:LR_longplay_bdd}
\end{figure}

Figure \ref{fig:LR_longplay_bdd} presents a comparison between a Java goal-driven test script and a BDD tactical-test scenario used to achieve game goals for the LabRecruits game. 
The LabRecruits-plugin has been extended with Cucumber, allowing Given-When-Then statements to be mapped to a wide range of controller and observer commands, as well as high-level tactics and goals. 
These BDD functions empower stakeholders to create and maintain human-readable game test scenarios without needing to deal with the underlying technical details of the controller and observer components in the plugin. 

\subsection{Long-play iv4XR-BDD test scenarios for Space Engineers}
\begin{figure}[!ht]
  \centering
  \includegraphics[width=\linewidth]{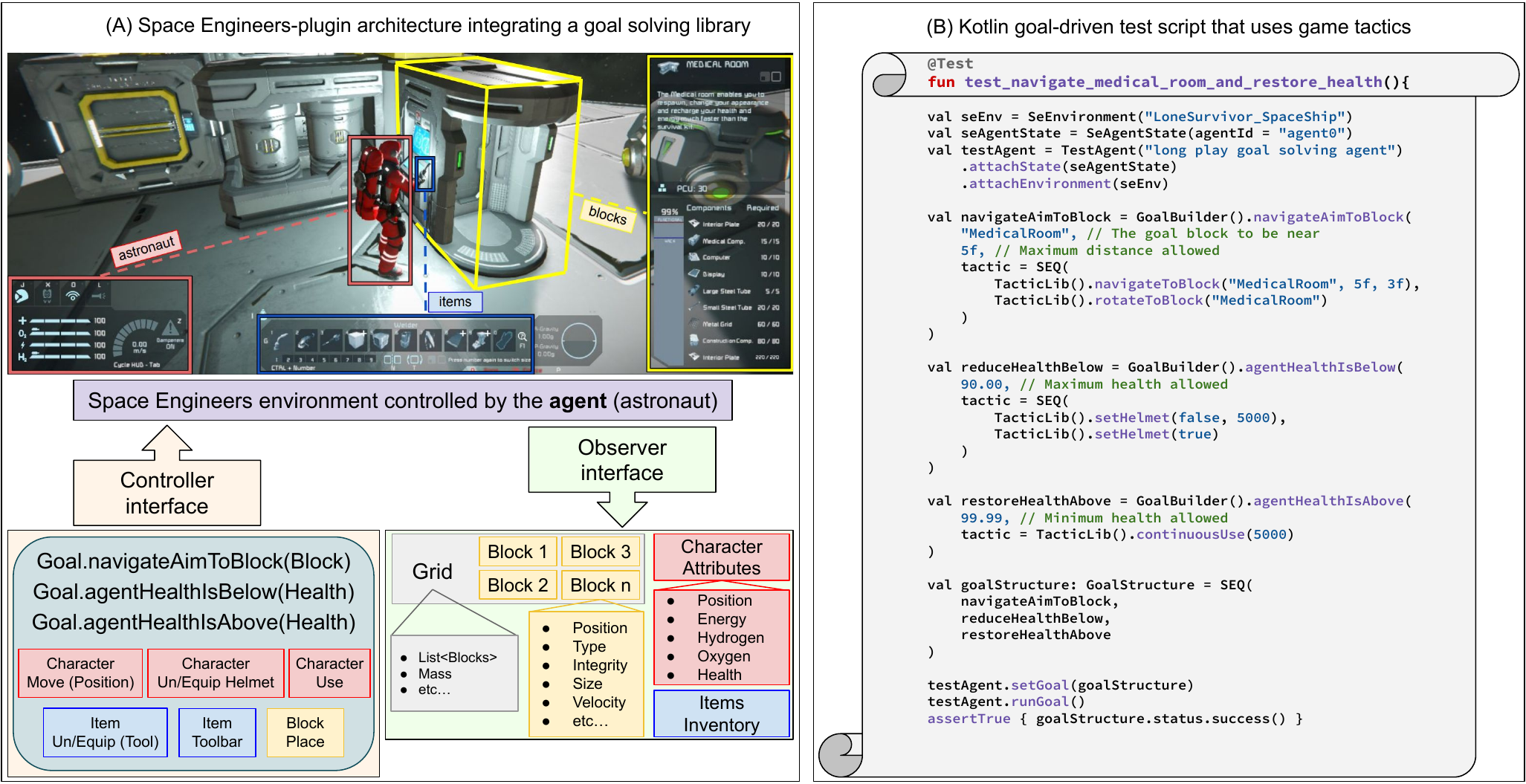}
  \caption{Overview of tactical programming in the Space Engineers-plugin with a Kotlin goal-driven test script that uses Space Engineers game tactics}
  \label{fig:SE_plugin_tactic}
\end{figure}

In order to support long-play test automation, we have extended the Space Engineers plugin by implementing the iv4XR agent's intelligent tactical capabilities. 
By integrating a high-level goal library with tactical programming, iv4XR agents can perform deliberative pathfinding and motion planning within 3D environments, allowing them to navigate complex game scenarios. 
Figure \ref{fig:SE_plugin_tactic} illustrates the incorporation of the high-level \textit{Goal} library within the controller interfaces of the Space Engineers-plugin architecture (A) and a Kotlin goal-driven test script that employs these tactics and goals to perform long-play tests in Space Engineers (B). 

In the Kotlin goal-driven test script example (B), the \textit{agent} is assigned three sub-goals: navigating and aiming toward the medical room block, temporarily removing its helmet to reduce health, and continuously using the medical room to restore health.
For each sub-goal, the \textit{agent} is instructed with Space Engineers tactics that guide its decision-making process to achieve these goals. 
These tactics internally employ game commands for movement, rotation, and interaction with game elements. 
Upon completion of these tactical executions, the \textit{agent} verifies that each sub-goal has been successfully accomplished. 

To continue supporting the creation, maintenance, and execution of human-readable test scenarios, we have expanded the BDD approach to create comprehensible tactical-test scenarios for long play-testing. 
This \ivxr{}-BDD approach enables mapping GWT statements into the tactical executions used to accomplish test goals. 
Additionally, this BDD solution allows \ivxr{} stakeholders to focus on higher-level abstract statements without considering the underlying distinctions between Java and Kotlin programming language test scripts. 

\begin{figure}[!ht]
  \centering
  \includegraphics[width=\linewidth]{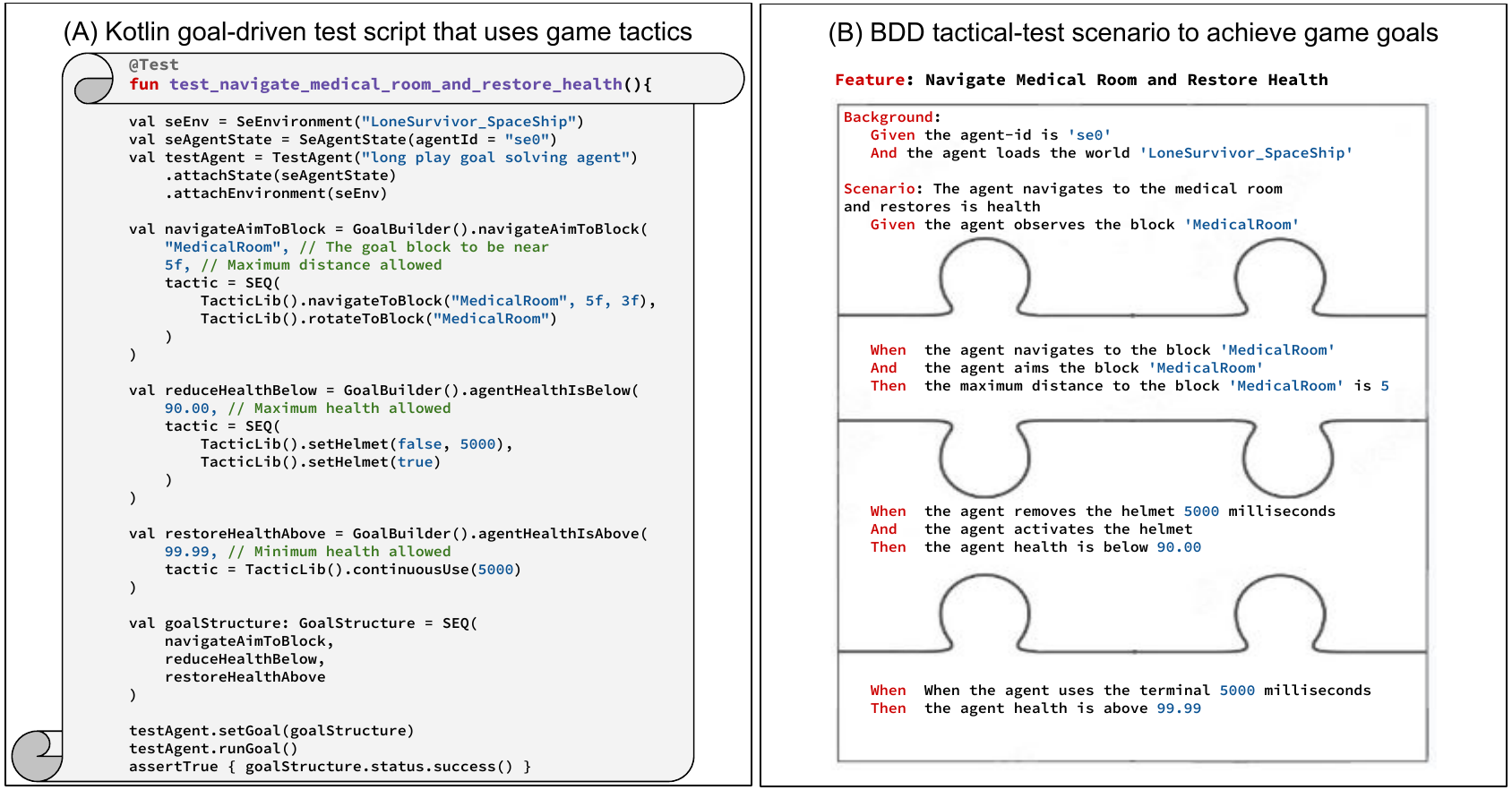}
  \caption{Space Engineers long-play iv4XR-BDD test scenario}
  \label{fig:SE_longplay_bdd}
\end{figure}

Figure \ref{fig:SE_longplay_bdd} shows a comparison between the Kotlin goal-driven test script and a BDD tactical-test scenario to perform long-play testing in the Space Engineers game. 
In this video, we can see how both implementations enable the same tactical-test automation to navigate to the medical room and test agent health restoration: \url{https://youtu.be/PidCHNzpExE} 

In this other video, the Kotlin goal-driven test script and the BDD tactical-test scenario are used to conduct long-play testing to progress in the Space Engineers blocks technology. 
The test focuses on navigating a survival scenario to obtain the necessary items to construct a Basic Assembler block completely: \url{https://youtu.be/5d1sH18uda4} 

\section{Conclusions and Future Work}
\label{section:conslusions.future.work}
This paper reports the industrial experience of developing a game plugin and adopting a BDD process to automate a subset of regression tests for the game Space Engineers. 
A BDD \textit{agent} has proven effectiveness and reliability in verifying the astronaut functionalities through natural language instructions, allowing the automation of 236 regression tests and showcasing its bug-detection capabilities. 
This automation is estimated to be equivalent to saving around 17 hours of manual effort per executed regression subset. 
These outcomes have encouraged the Space Engineers team to develop a dedicated server for seamless integration of automated regression testing into nightly builds during the 3-4 month development cycle periods.  
We hope this experience motivates and guides other stakeholders in integrating a framework to map internal entity observation and game action execution with natural language statements to enable BDD testing. 

Motivated by the success of this BDD implementation, the \ivxr{} project team extended the \ivxr{} framework by incorporating BDD tactical \textit{agents} that facilitate long-play testing of levels in both the experimental 3D game LabRecruits and Space Engineers. 
For LabRecruits, this BDD integration allows non-technical stakeholders to learn and understand the tactical and goal-solving capabilities of the \ivxr{} framework. 
In the case of Space Engineers, the new tactical capabilities of the \ivxr{}-BDD \textit{agents} support ongoing test automation efforts, enabling long test runs that simulate complex user interactions and navigational decision-making within game scenarios. 

In future work, the Space Engineers team will continue automating the remaining regression test scenarios and extending the regression test suite to validate the placement and properties of blocks. 
Additionally, we are working to overcome challenges such as 
(i) determine how internal game observations may differ from the user's GUI perspective (e.g., users must scroll the inventory to find an item), 
(ii) research approaches to integrate reliable graphics and sound oracles within the \ivxr{} Space Engineers-plugin. 

Regarding the capabilities of the \ivxr{} framework, we plan to 
(iii) investigate the integration of the framework with popular game engine platforms such as Godot and Unity. This integration is expected to streamline the observation of internal game objects and facilitate the development of game plugins. 
(iv) As part of these enhancements, qualitative and quantitative empirical studies will be conducted with the Space Engineers team and stakeholders from other games adopting the \ivxr{} framework. In particular, to promote the use of the \ivxr{}-BDD solution, we want to continue evaluating the effectiveness and intention to use the approach. 
Finally, (v) we plan to explore the benefits of BDD practices in other interactive system domains, such as surgical systems, autonomous vehicles, or immersive education systems. 

\section*{Acknowledgement}
This work has been partially funded by: H2020 EU iv4XR grant nr. 856716 and ENACTEST ERASMUS+ grant nr. 101055874. 

\section*{Statement}
During the preparation of this work the author(s) used ChatGPT and Grammarly in order to improve the grammar and syntax of the content. 
After using this tool/service, the author(s) reviewed and edited the content as needed and take(s) full responsibility for the content of the publication. 

\bibliographystyle{elsarticle-harv}
\bibliography{testar.bib}

\end{document}